\documentstyle[prl,aps]{revtex}
\newcommand{\BE}{\begin{equation}}
\newcommand{\EE}{\end{equation}}
\newcommand{\BA}{\begin{eqnarray}}
\newcommand{\EA}{\end{eqnarray}}

\begin{document}

\twocolumn[\hsize\textwidth\columnwidth\hsize\csname@twocolumnfalse\endcsname
\title{Localization of acoustic waves in
1D random liquid media}
\author{Pi-Gang Luan and Zhen Ye}
\address{Department of Physics and Center for Complex Systems,
National Central University, Chung-li, Taiwan, ROC}

\date{\today}

\maketitle
\begin{abstract}
We study acoustic propagation in one dimensional water ducts
containing many air-filled blocks. The acoustic band structures
for the periodic arrangements of the blocks is calculated, whereas
the transmission for various random configurations of the blocks
is computed by the transfer matrix method. The results show that
while all waves are localized for any given amount of disorders,
there is no genuine scaling behavior for the system. The results
also reveal a distinct collective behaviour for localized waves, a
feature useful for distinguishing the localization from the
residual absorption effect. \vspace{4pt}
\end{abstract}
 ]

The fact that the electronic localization in disordered
systems\cite{Anderson} is of wave nature has led to suggestion
that classical waves could be similarly localized in random
systems. The effort in searching for localization of classical
waves such as acoustic and electro-magnetic waves is tremendous.
It has drawn intensive attentions from both theorists
\cite{Frisch73,Hodges82,Baluni,SEGC,Sor,Maradudin93,Ye1,Ye2,SIAM}
and experimentalists\cite{Hodges83,Sch}.

Localization of waves in one dimensional (1D) systems has
attracted particular interest from scientists because in higher
dimensions the interaction between waves and scatterers is so
complicated that the theoretical computation is rather involved
and most solutions require a series of approximations which are
not always justified, making it difficult to relate theoretical
predictions to experimental observations. Yet wave localization in
one dimension (1D) poses a more manageable problem which can be
tackled in an exact manner by the transfer matrix method.
Moreover, results from 1D can provide insight to the problem of
wave localization in general and are suitable for testing various
ideas. Indeed, over the past decades considerable progress has
been made in understanding the localization behavior in 1D
disordered systems\cite{Lifshits}. However, a number of important
issues remained untouched. These issues include, for example, how
waves are localized inside the media and whether there is a
distinct feature for wave localization which would allow to
differentiate the localization from residual absorption effect
without ambiguity\cite{Frank,Cha}. Results from the statistical
analysis of the scaling behavior in 1D random media is not
conclusive. A further question could be whether the localized
state is a phase state which would accommodate a more systematic
interpretation from the view of a symmetry breaking and collective
behavior, in analogy to phase states such as superconductivity.
This Letter attempts to provide insight to these questions.

Here we study the problem of acoustic wave propagation in one
dimensional water ducts containing many air blocks either
regularly or randomly but on average regularly distributed inside
the ducts. The frequency band structures and wave transmission are
computed numerically. We show that while our results affirm the
previous claim that all waves are localized inside an 1D medium
with any amount of disorder, there are, however, a few distinctive
features in our results. Among them, in contrast to optical
cases\cite{Deych}, there is no universal scaling behavior in the
present system. In addition, when waves are localized, a
collective behaviour of the system emerges.

Assume that $N$ air blocks of identical thickness $a$ are placed
regularly or randomly in a water duct with length $L$ measured
from the left boundary of the duct (LB). The air fraction is
$\beta = Na/L$, the average distance between two adjacent water
layers is $\langle d\rangle = L/N=a/\beta$, and the average
thickness of water layers is $\langle
b\rangle=(\sum^{N}_{j=1}b_{j})/N$. The degree of randomness for
the system is controlled by a parameter $\Delta$ in such a way
that the thickness of the $j$-th water layer is $b_{j}=\langle
b\rangle(1+\delta_{j})$ with $\delta_{j}$ being a random number
within the interval $[-\Delta,\Delta]$; the regular case
corresponds to $\Delta =0$. An acoustic source placed at LB
generates monochromatic waves with an oscillation
$v(t)=ve^{-i\omega t}$. Transmitted waves propagate through the
$N$ air layers and travel to the right infinity. In order to avoid
unnecessary confusion, possible effects from surface tension,
viscosity or any absorption are neglected. For convenience, we use
the dimensionless quantity $k\langle b\rangle$ to measure the
frequency, where $k=\omega/c$ is the wave number and $c$ is the
sound speed in water. Similarly, $k_{g}$ and $c_{g}$ represent the
wave number and sound speed in the air blocks respectively.

The wave propagation in such a system can be solved using the
transfer matrix method\cite{Baluni}. After dropping out the time
factor $e^{-i\omega t}$, the wave propagation is governed by two
equations\cite{Ish}. The first is the Helmholtz equation, \BE
p_{m}''(x)+k^{2}_{m} p_{m}(x)=0,\label{helm}\EE in which
$p_{m}(x)$ is the pressure field, and the subscript $m$ refers to
the medium that can be either water or air, depending on where $x$
is located. Within any layer, Eq.~(\ref{helm}) warrants two
solutions: $A_{m}e^{ik_mx}$ represents the wave transmitted away
from the source to the right and $B_{m}e^{-ik_mx}$ the wave
reflected towards the source. The total wave is therefore
$p_{m}(x)=A_{m}e^{ik_mx}+B_{m}e^{-ik_mx}$. The second equation
relates the oscillation velocity and the pressure field, \BE
{u}_{m}(x) =\frac{1}{i\omega\rho_{m}} p'_{m}(x)=
(A_{m}e^{ik_mx}-B_{m}e^{-ik_mx})/\rho_m c_m,\label{pu}\EE where
$\rho_{m}$ refers to the equilibrium mass density of medium $m$.

The coefficients $A_m$ and $B_m$ in any two adjacent layers are
connected by a transfer matrix. By invoking the condition that the
pressure and velocity fields are continuous across the interfaces
separating water and air and the condition that there is no
reflected wave to the right end of the system, the matrix elements
for the transfer matrices linking all interfaces can be deduced,
and waves in any particular block is therefore completely
determined. The ratio between the amplitude of the outgoing wave
at the right boundary and that of the transmitted wave at the
source defines the transmission rate.

Before going further, a general discussion on wave propagation is
in place. When waves propagate through media alternated with
different material compositions, multiple scattering of waves is
established by an infinite recursive pattern of rescattering. In
terms of wave fields, the energy flow in the system is calculated
from $J \sim \mbox{Re}[i (p^{*}(x)\partial_{x} p(x)]$. Writing
$p(x)=|A(x)|e^{i\theta(x)}$ with $|A|$ and $\theta$ being the
amplitude and phase respectively, the energy flow becomes $J \sim
|A|^{2}\partial_{x}{\theta}$. Obviously, the energy flow will come
to a complete halt and the waves could be localized in space when
phase $\theta$ is constant and $|A|$ does not equal zero.

First consider the case with the periodic placement of air blocks
(with spatial period $d$, water layer thickness $b$). According to
Bloch theorem, and the wave field $p$ can be written as \BE
p(x)=A(x)e^{iKx}, \EE where $A(x)$ is a periodic function
satisfying $A(x+d)=A(x)$, and $K$ is the Bloch wave number
represented in the dispersion relation \BE \cos Kd=\cos k_{g}a\,
\cos kb -\cosh 2\eta\,\sin k_{g}a\, \sin kb.\label{disp}\EE Here
$k_g = k/h$ and $\eta=\ln q$ with $q^{2}=gh$ and
$g=\rho_{g}/\rho$, $h=c_{g}/c$.

\begin{figure}[hbt]
\epsfxsize=3in\epsffile{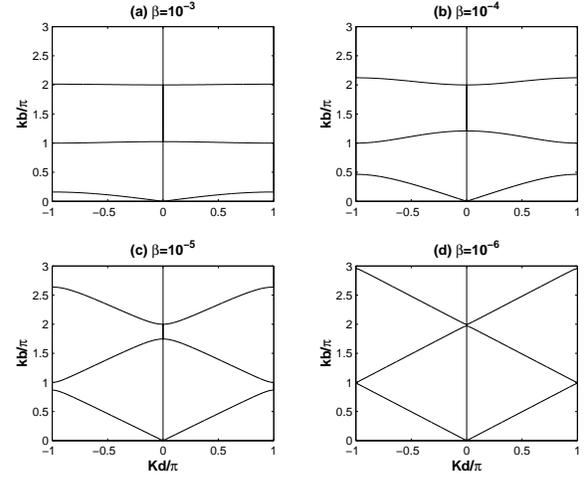}
\vspace{10pt}\caption{\label{figure1}\small Band structures of an
infinite periodic system for four different air fractions
($\beta$). In the plots, $K$ is the Bloch wave number and $d$ is
the lattice spacing.}
\end{figure}

The band structures for four air-fractions are shown in
Fig.~\ref{figure1}. The ranges covered by the dispersion curves
refer to the passing bands, while the areas sandwiched by any two
curves to the complete band gaps. Within the gap, waves are
evanescent and cannot propagate. Since the air blocks are strong
scatterers due to the large contrast in acoustic impedance between
air and water, we see that even a small air-fraction can lead to
wide band gaps as seen in Fig.~\ref{figure1}(a). With increasing
$\beta$, the passing bands are narrowed and become streaks. When
the air-fraction is reduced, however, the band gaps gradually
disappear, as shown by Fig.~\ref{figure1}(d).

\input{epsf}

\begin{figure}[tbh]
\epsfxsize=3in\epsffile{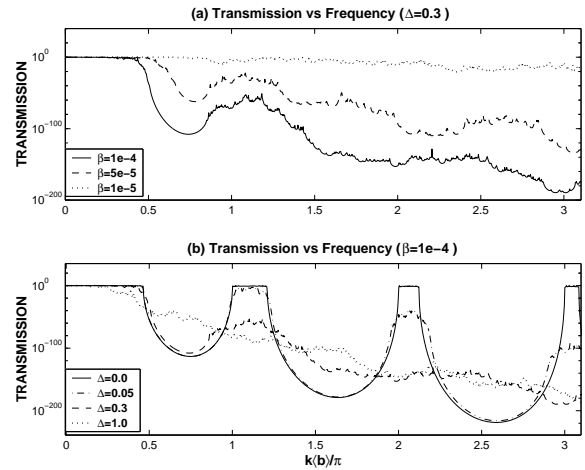}
\vspace{10pt}\caption{\label{figure2}\small Transmission versus
$k\langle b\rangle/\pi$ for various air fractions at $\Delta =
0.3$ (a) and different disorders (b). The number of the air-blocks
is 100.}
\end{figure}

Wave propagation properties can be significantly affected by
varying air-fraction or adding randomness. Fig.~\ref{figure2}(a)
presents the typical results of the transmission rate as a
function of $k\langle b\rangle$ for various $\beta$ at a given
randomness.  At frequencies for which the wavelength is smaller
than the averaged distance between air blocks, the transmission is
significantly reduced by increasing air fraction.

\input{epsf}

Fig.~\ref{figure2}(b) illustrates the effect of the randomness
$\Delta$ on transmission for a given air-fraction. For comparison,
the transmission in the corresponding regular array ($\Delta = 0$)
is also plotted. The gaps are located between $k\langle
b\rangle/\pi = 0.46$ and $1$, $1.21$ and $2$, $2.12$ and $3$, and
so on. We find that for frequencies located inside the band gaps
of the corresponding regular array, the disorder-induced
localization effect competes yet reduces the band gap effect. To
characterize wave localization in this case, both the band gap and
the disorder effects should be considered, supporting the two
parameter scaling theory\cite{Deych}. However, increasing disorder
tends to smear out the band structures. When exceeding a certain
amount, the effect from the disorder suppresses the band gap
effect completely, and there is no distinction between the
localization at frequencies within and outside the band gaps.

Fig.~\ref{figure2} shows that the localization behaviour depends
crucially on whether the wavelength ($\lambda$) is greater than
the average distance between air blocks. When $\langle
b\rangle/\lambda$ is less than one, the localization effect is
weak. We also observe that with the added disorder, the
transmission is enhanced in the middle of the gaps. Similar
enhancement due to disorder has also been reported
recently\cite{Maradudin95}. Differing from \cite{Maradudin95},
however, the transmission at frequencies within the gaps of the
corresponding periodic arrays in the present system is not always
enhanced by disorder. Instead the transmission is reduced further
by the disorder near the band edges.

\begin{figure}[tbh]
\epsfxsize=3in\epsffile{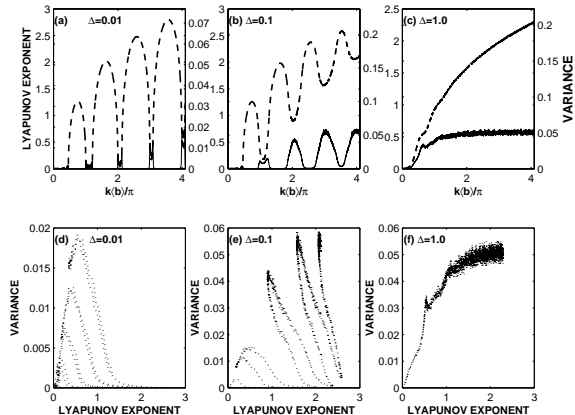}
\vspace{10pt}\caption{\label{figure3}\small Diagrams (a), (b), and
(c) show the Lyapunov exponent (LE) in broken lines and its
variance in solid lines as a function of $k\langle b\rangle/\pi$.
Diagrams (d), (e), and (f) present the plots of the exponent
versus its variance. Here $\beta=10^{-4}$.}
\end{figure}

Fig.~\ref{figure3} presents the results for Lyapunov exponent (LE)
and its variance as a function of non-dimensional frequency
$k\langle b\rangle /\pi$. At low disorders, the variance of LE
inside the gaps is small. Contrast to the optical
case\cite{Deych}, there are no double maxima inside the gap. With
increasing disorder, double peaks appears inside the allow bands.
When exceeding a certain critical value, however, the double peaks
emerge. The higher frequency, the lower is the critical value. For
example, the double peaks are still visible in the first allow
band (c.~f. Fig.~\ref{figure3}(b)), while there is only one peak
inside the higher passing bands. Meanwhile, the increasing
disorder reduces the band gap effect and smears LE, in accordance
with Fig.~\ref{figure2}. We also plot LE-variance relation in
Fig.~\ref{figure3}. With increasing disorder, we do not observe
genuine linear dependence between LE and its variance, as expected
from the single parameter scaling theory.

In the past the localization phenomenon in 1D is usually
characterized by LE, here we propose to use a phase behavior of
the waves to characterize the localization. We compute the energy
density inside the sample from \BE
E_{m}(x)=\frac{\rho_{m}}{4}\left(|u_{m}|^{2}+\frac{|p_{m}|^{2}}
{\rho^{2}_{m}c^{2}_{m}}\right),\label{E}\EE where $m$ refers to
either water or air. Meanwhile, the phases of $p(x)$ and $u(x)$
are recorded. Expressing $p(x)$ and $u(x)$ as
$A_{p}(x)e^{i\theta_{p}(x)}$ and $A_{u}(x)e^{i\theta_{u}(x)}$, we
construct unit phase vectors
$\vec{v}_{p}=\cos\theta_{p}\hat{e}_{x}+\sin\theta_{p}\hat{e}_{y}$
and
$\vec{v}_{u}=\cos\theta_{u}\hat{e}_{x}+\sin\theta_{u}\hat{e}_{y}$.
Then the behavior of the phase vectors along the path are
investigated. Physically, these phase vectors represent the
oscillation behavior of the system.

\begin{figure}[tbh]
\epsfxsize=3.2in\epsffile{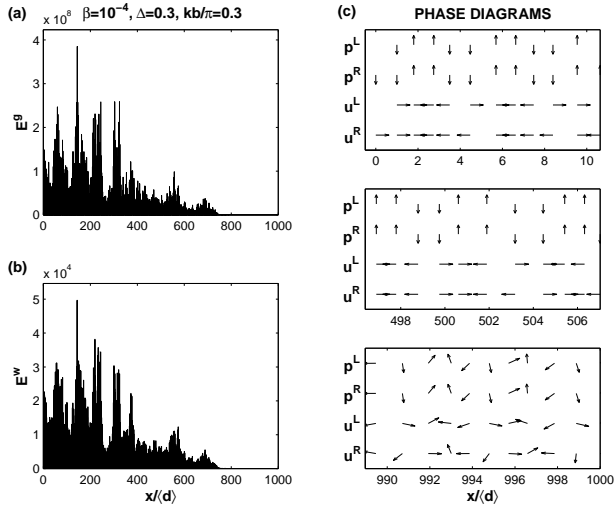}\vspace{10pt}
\caption{\label{figure4}\small Energy distribution and phase
diagrams for localized state.  Here $\beta=10^{-4}$, $\Delta=0.3$
and $k\langle b\rangle=0.3\pi$ is a frequency in the first allowed
band. (a) Energy densities in the air blocks. (b) Energy densities
in water. (c) Phase vectors at the interfaces for three spatial
ranges of the medium. The unit of energy density is $J/m^{3}$.}
\end{figure}

Typical results for the spatial distribution of the energy and the
phase behaviour for the given disorder and air fraction are shown
in Fig.~\ref{figure4}. Here for the sake of convenience, only the
phase vectors at the interfaces between air and water are shown.
First, we note that the energy density is constant in each
individual block. This is a special feature of 1D classical
systems, and can be verified by a deduction from Eq.~(\ref{E}). We
find that when the sample size is sufficiently large, waves are
always localized for any given amount of randomness. When
localized, the waves are trapped inside the medium, but not
necessarily confined at the site of the source, unless the band
gap effect is dominant. The energy distribution does not follow an
exponential decay along the path. This differs from situations in
higher dimensions\cite{Ye1,Ye}. It is also shown that the energy
stored in the medium can be tremendous (c.~f.
Figs.~\ref{figure4}(a) and (b)). With increasing sample size, the
peak amplitude may grow, pointing to the stochastic resonance, in
agreement with \cite{Frisch73}. When disorder is weak and for
frequencies within the gaps, such a stochastic resonance behavior
disappears.

It is also observed that for all frequencies, there is a
collective behavior for the phase vectors. In
Fig.~\ref{figure4}(c) we plot the phase vectors in three spatial
regimes, namely, near the transmitting site, in the middle of the
duct, and at the far end from the source. Symbols $p^{L}$,
$p^{R}$, $u^{L}$, $u^{R}$ appearing in Fig.~\ref{figure4}(c)
denote respectively the phase vectors for the pressure and the
velocity fields on the left and right side of the air blocks. It
is clear that when waves are localized, all the phase vectors of
the pressure field are pointing to either $\pi/2$ or $-\pi/2$, and
perpendicular to the phase vectors of the velocity field. The
pressure at the two sides of any air block varies in phase.
Mostly, the two sides also oscillate in phase. Different from
higher dimensional cases in which all phase vectors of localized
fields point to the same direction\cite{Ye2,Ye}, the present phase
vectors are constant by domains; this ensures no energy flows. The
velocity field in neighboring domains oscillate exactly out of
phase. The phase vector domains are sensitive to the arrangement
of the air blocks. We stress that such a phase ordering not only
exists for the boundaries of the air blocks, but also appears
inside the whole medium.

\input{epsf}

The coherence behavior is a unique feature for wave localization,
which could be verified by the cross correlation measurement. At
the far end of the sample, however, the phase vectors become
gradually disoriented, implying that the energy can leak out only
at the boundary due to the finite sample size. This boundary
effect vanishes exponentially as the sample size is increased. The
fact that the phase vectors are constant in domains indicates that
once waves are localized, no more energy can be pumped into the
system. When localization is evident, increasing the sample size
by adding more air blocks will not change the patterns of the
energy distribution and phase vectors. Therefore the energy
localization and the phase behavior are not caused by the boundary
effect.

In summary, we have demonstrated a phase transition in acoustic
propagation in an 1D random liquid medium. The results have shown
that waves are always confined in a finite spatial region. The
disorder leads a significant energy storage in the system. It is
also indicated that the wave localization is related to a
collective behavior of the system in the presence of multiple
scattering, also observed for higher dimensions\cite{Ye2,Ye}. The
appearance of such a collective phenomenon may be regarded as an
indication of a kind of Goldstone modes in the context of the
field theory\cite{Umezawa}.

The work received support from National Science Council (No.
NSC89-2611-M008-002 and NSC89-2112-M008-008).

\end{document}